\documentclass[amsmath,amssymb,preprint,12pt]{revtex4}

\begin{document}

\title{New Hints from General Relativity\footnote{This %
essay received an ``honorable mention" in the 2003 Essay %
Competition of the Gravity Research Foundation}}
\date{\today}
    \author{Olaf Dreyer}
    \email{odreyer@perimeterinstitute.ca}
    \affiliation{Perimeter Institute for Theoretical Physics, 35 King
    Street North, Waterloo, Ontario N2J 2W9, Canada}

\begin{abstract}
The search for a quantum theory of gravity has
followed two parallel but different paths. One aims at
arriving at the final theory starting from a priori assumptions as to its form
and building it from the ground up. The other
tries to infer as much as possible about the unknown theory
from the existing ones and use our current knowledge to constrain
the possibilities for the quantum theory of gravity.

Probably the biggest success of the second path has been the
results of black hole thermodynamics. The subject of this essay is
a new, highly promising such result, the application of
quasinormal modes in quantum gravity.
\end{abstract}

\maketitle

\subsection{Introduction}
For almost sixty years, a major goal of theoretical
physics has been to write an article on quantum gravity that does not start
by recounting how long people have been trying with limited
success. In their efforts, researchers in the field have followed two different paths.
The first involves starting from basic principles and  constructing
a theory from the ground up. The second tries to accumulate as
many hints as possible from the known and established theories, in
the hope that this knowledge will constrain the possibilities for a quantum theory of gravity.

The best known examples of the first kind of approach are String
Theory (see \cite{pol} for a recent introduction to the subject)
and Loop Quantum Gravity (see \cite{Thiemann:2001yy} for a
comprehensive overview). Both of these use
conventional quantum theory. String theory is the quantum theory
of string-like objects in space-time, whereas Loop Quantum Gravity
is the attempt to quantize the gravitational field itself starting from
the classical field equations.

By far the most important results of the second kind are
Bekenstein's and Hawking's results on black hole thermodynamics.
Starting from theorems in classical general relativity, Bekenstein
\cite{b:ent} argued that a black hole has a temperature $T$ which
is proportional to its surface gravity $\kappa$ and an entropy $S$
that is proportional to its area $A$. Using quantum field theory
on curved space-time, Hawking \cite{Hawking:1975sw} was then able to
fix the proportionality constants and establish the relations
\begin{equation}\label{eqn:temp}
  T = \frac{\kappa\hbar}{2\pi}
\end{equation}
and
\begin{equation}\label{eqn:ent}
   S = \frac{A}{4 \hbar G}.
\end{equation}
In this way, well-established theories,  namely, general relativity and
quantum field theory,  give us important insights into the
problem of quantum gravity. The insight we should keep from this is that,
given a classical black hole with area $A$, there should be
\begin{equation}\label{eqn:number}
  \exp(A/4)
\end{equation}
quantum mechanical states which, to a macroscopic observer, look
like the given black hole. The important question then is: {\em
What are these states?} It is arguably the most important success
of String Theory and Loop Quantum Gravity so far that they can
provide partial answers to this question.

In the next section we will discuss how Loop Quantum Gravity
achieves this. In the process, we will encounter a new and
surprising hint from General Relativity which provides additional
insight into the problem, and which goes beyond the particular
approach of Loop Quantum Gravity. We will close with a comparison
of the two paths and an outlook.

\subsection{Counting the Black Hole States... Almost}
A basis for the Hilbert space of Loop Quantum Gravity is given by
spin networks. These are graphs whose edges are labelled by
representations of the gauge group of the theory. In the case of
gravity this group is taken to be SU(2) and the representations
are thus labelled by positive half-integers $j = 0, 1/2, 1, 3/2,
\ldots$. If a surface is intersected by an edge of the spin
network carrying the label $j$, the surface acquires the area
\cite{rovsmo, ashlew}
\begin{equation}\label{eqn:area}
  A(j) = 8\pi l_P^2 \gamma \sqrt{j(j+1)},
\end{equation}
where $l_P$ is the Planck length and $\gamma$ is the so-called
Immirzi parameter \cite{Immirzi:1996dr}. It parameterizes an
ambiguity in the choice of canonically conjugate variables that
are used in the quantization. There is no a priori reason to fix
this parameter to any particular value.

One can think of the area of a black hole as being a
consequence of a large number of spin network edges puncturing its
surface (see \cite{smolin, rovelli, krasnov, abck, abk}). Each
edge with spin $j$ contributes the amount of area given by formula
(\ref{eqn:area}) to the black hole area. On the horizon, such a puncture
also increases the dimensionality of the Hilbert space of the
theory living on the boundary. Each puncture of an edge with spin
$j$ increases the dimension by a factor of $2j +1$, namely, by the
dimension of the spin $j$ representation. If there is a large
number $N$ of edges with spins $j_i$, $i=1, \ldots, N$,
intersecting the horizon the dimension of the boundary Hilbert
space is
\begin{equation}
  \prod_{i=1}^N (2j_i + 1).
\end{equation}

The entropy of a black hole with a given area $A$ is simply given by
the logarithm of the dimension of the Hilbert space of the
boundary theory. It can be shown that the statistically most
important contribution comes from those configurations in which
the lowest possible spin dominates. Let us denote this spin by
$j_{\min}$. The entropy then is
\begin{equation}\label{eqn:entropy}
    S = N \ln (2j_{\min}+1),
\end{equation}
where $N$ can be calculated from the area $A$ of the black hole
and from the amount of area $A(j_{\min})$ contributed by every
puncture. One obtains
\begin{equation}\label{eqn:N}
  N = \frac{A}{8\pi l_P^2\gamma\sqrt{j_{\min}(j_{\min}+1)}}.
\end{equation}
We already see that the entropy of the black
hole turns out to be proportional to the area, as we had hoped.
The fly in the ointment is that we can not
fix the constant $\gamma$.

We will again turn to general relativity for help. It comes from
the study of the so-called quasinormal modes of a black hole.

\subsection{A New Hint: Quasinormal Modes in Quantum Gravity}

The reaction of a black hole to perturbations is dominated by a
discrete set of damped oscillations called quasinormal modes (see
\cite{nollertrev, kokkotas} for more details.) A remarkable
property of these modes is that in the large damping limit the
real part of their frequencies $\omega$ becomes a non-zero
constant. It was shown in \cite{lubos1, lubos2} that this real
part is equal to
\begin{equation}\label{eqn:realomega}
  \frac{\ln 3}{8 \pi M}.
\end{equation}

Thus, a particular frequency is associated to any black hole.
If we had a quantum theory of black holes, we would
expect that there is a transition between two black hole states
that gives rise to exactly this classical frequency. In the
picture of a black hole described in the previous section there is a natural
candidate for such a transition. It is the appearance or
disappearance of a puncture with spin $j_{\min}$. This causes a change in
the area of the
black hole  given by
equation ({\ref{eqn:area}):
\begin{equation}\label{eqn:deltaa}
    \Delta A = A(j_{\min}) = 8 \pi l_P^2 \gamma \sqrt{j_{\min}(j_{\min}+1)}
\end{equation}

We can now fix the Immirzi parameter $\gamma$ appearing in this
equation simply by requiring that the change $\Delta M$ in the mass,
corresponding to this change in area, equals the energy of a
quantum with the frequency of equation (\ref{eqn:realomega}).  That is,
we fix $\gamma$ by setting
\begin{equation}\label{eqn:equal}
    \Delta M = \frac{\hbar\ln 3}{8 \pi M}.
\end{equation}
Since the area $A$ and the mass $M$ of a Schwarzschild black hole
are related by $A = 16 \pi M^2$ the mass change of equation
(\ref{eqn:equal}) translates into the area change
\begin{equation}\label{eqn:areaj}
  \Delta A = 4\ln3\; l_P^2.
\end{equation}
Comparing this with equation (\ref{eqn:deltaa}) gives the desired
expression for the Immirzi parameter $\gamma$:
\begin{equation}\label{eqn:newgam}
  \gamma = {\frac{\ln 3}{2\pi\sqrt{j_{\min}(j_{\min}+1)}}}.
\end{equation}
Using this value for the Immirzi parameter in formulae
(\ref{eqn:entropy}) and (\ref{eqn:N}) gives
\begin{equation}\label{eqn:sfinal}
  S = \frac{A}{4 l_P^2}\frac{\ln (2j_{\min} +1)}{\ln 3}.
\end{equation}
We thus see that we get an exact agreement with the Bekenstein -
Hawking result if we set $j_{\min}=1$.

\subsection{Discussion}
The classical theory of general
relativity is still able to give us hints about the so far elusive
quantum theory of gravity. Using the unique feature of the
quasinormal mode spectrum from the classical theory,
we were able to argue that a free
parameter appearing in a \emph{quantum} theory should be fixed to
a certain value. What is even more amazing is that this gives
the right value for the entropy of a black hole.

But it does not stop here. To get the right value for the black
hole entropy we had to choose $j_{\min} = 1$. It was argued in
\cite{olaf} that this implies that the appropriate gauge group to
be used in quantum gravity is SO(3) instead of SU(2). Another
surprising insight on the quantum theory, resulting from a closer
look at the classical theory.

The results mentioned in the previous section are just one example
of how useful the classical theory can be in our search for a
quantum theory of gravity. We wish to mention one more recent such
example. Using an extended notion of symmetry, one can argue that a
generic black hole spacetime possesses in the near horizon region
a set of symmetry vector fields $\xi_n$, $n\in\mathbb{N}$, that
form a Diff($S^1$) (see \cite{carlip} and \cite{dgw} for more
details). When represented on phase space, the Diff($S^1$) acquires
a central charge and we thus obtain a Virasoro algebra. Knowing
the representation theory of the Virasoro algebra
then gives the dimensionality of the corresponding representation
space. The logarithm of this dimension turns out to be equal to
the Bekenstein - Hawking entropy of a black hole.

This leads to another hint from general relativity. The symmetry
vector fields found in the \emph{classical} black hole space-time
tell us about the space of states of the \emph{quantum} theory.
Given a quantum theory of black holes we should look for
representations of a Virasoro algebra in it.

One might ask: Why bother? Since we have the final theory already,
string theory or loop quantum gravity,
or at least we are very close to completing it, aren't these kinds of
investigations just distractions from the real work?

I would argue that this is not so. To this day, it is far from
clear whether string theory or loop quantum gravity is the right
quantum theory of gravity.  There are plenty of open questions.
Both in the way they treat black holes, and in
their general status.

String theory is not yet able to deal with non-extremal black holes
since all the tools so far rely heavily on supersymmetry.
The loop quantum gravity treatment of black hole entropy
requires the a priori assumption of a horizon surface. One should
instead be able to infer the existence of such a surface from the
theory itself.

On a more general level, string theory holds the promise
of a beautiful unified theory but, for the moment, suffers from an
embarrassment of riches. The number of possible vacua seems to be
so large (\cite{suskind}) that is is not clear what the predictive
content of the theory is. Loop quantum gravity has shown us
how to quantize a background independent theory, but has yet to
convincingly argue how general relativity is recovered
in the large scale limit.

The results we have been discussing in this essay are of a
different nature. They are likely to be true in any theory. The
example we have been looking at in some detail might be
interpreted as saying that a quantum theory describing a black
hole should allow for spin one excitations. This can be true
even outside the framework of loop quantum gravity.

As long as we do not have the final theory,  results of
the kind discussed here are still very important and we can only
hope that general relativity has some more hints in store for us.

\end{document}